\documentclass[11pt,twoside,a4paper]{article}
\usepackage{epsfig}
\usepackage{graphicx}
\usepackage{srt_style}
\begin{document}
\def\lsim{\, \lower2truept\hbox{${< \atop\hbox{\raise4truept\hbox{$\sim$}}}$}\,}
\def\gsim{\, \lower2truept\hbox{${> \atop\hbox{\raise4truept\hbox{$\sim$}}}$}\,}
%
\title{Millimeter-band Surveys of Extragalactic Sources}

\author{Gianfranco De Zotti, Francesca Perrotta and Gian Luigi Granato}
\affil{INAF -- Osservatorio Astronomico di Padova, Italy}
\author{Laura Silva}
\affil{INAF -- Osservatorio Astronomico di Trieste, Italy}
\author{Roberto Ricci, Carlo Baccigalupi and Luigi Danese}
\affil{SISSA, Trieste, Italy}
\author{Luigi Toffolatti}
\affil{Departamento de F{\'\i}sica, Universidad de Oviedo, Spain}
%
\begin{abstract}
Surveys at mm wavelengths emphasize rare classes of extragalactic
radio sources characterized by spectra keeping flat or inverted up
to high frequencies (such as blazars, GPS sources, advection
dominated sources, or even gamma-ray burst afterglows), which are
are difficult to single out at lower frequencies where the counts
are dominated by far more numerous populations. Such surveys may
lead to the discovery of populations with extreme properties, such
as sources of the general GPS or blazar type but peaking at mm
wavelengths, whose existence is hinted by currently available
data. At relatively lower flux levels mm counts are dominated by
high-redshift dusty galaxies, such as those detected by SCUBA and
MAMBO surveys. These data have profound implications for our
understanding of the formation and early evolution of galaxies.
High sensitivity surveys at cm/mm wavelengths may also detect
galaxy-scale Sunyaev-Zeldovich effects due to gas heated by
central active nuclei.
\end{abstract}

\section{Introduction}

The millimeter region is quite a special one, since its
corresponds to a minimum in the spectral energy distribution of
the Galaxy and of most (but not all) classes of extragalactic
sources. The minimum occurs at the cross-over between radio
emission, which generally (but not always) decreases with
frequency with a power law spectrum ($S_\nu \propto
\nu^{-\alpha_r}$, with $\alpha_r \sim 0.5$ to $1$), and dust
emission which steeply rises with frequency (with a spectral index
$\alpha_d \sim -3$ to $-4$). Just because of such steep rise the
frequency of the minimum is only weakly dependent on the relative
intensity of the two components. Moreover, the effect of redshift
on the dust emission peak is to some extent compensated by the
increase in dust temperature associated with the luminosity
evolution of sources. This is why this spectral band is optimal
for mapping the Cosmic Microwave Background. For the same reason,
it is also optimal for singling out the rare, but very
interesting, source populations with non-standard spectra, which
at other frequencies are submerged by the much more numerous
populations with ``standard'' spectra.

In particular, the high-frequency selection allows us to beat the
strong, free-free or synchrotron, self-absorption occurring in the
densest emitting regions which are normally the closest to the
central engine. As discussed in the following, this allows us to
investigate both the earliest and the late phases of nuclear
activity. But the millimeter region also carries key information
on more conventional classes of radio sources and allows us to
investigate the early phases of dust-obscured galaxy evolution.

\begin{figure}
\includegraphics[width=15cm,height=14cm,angle=0]{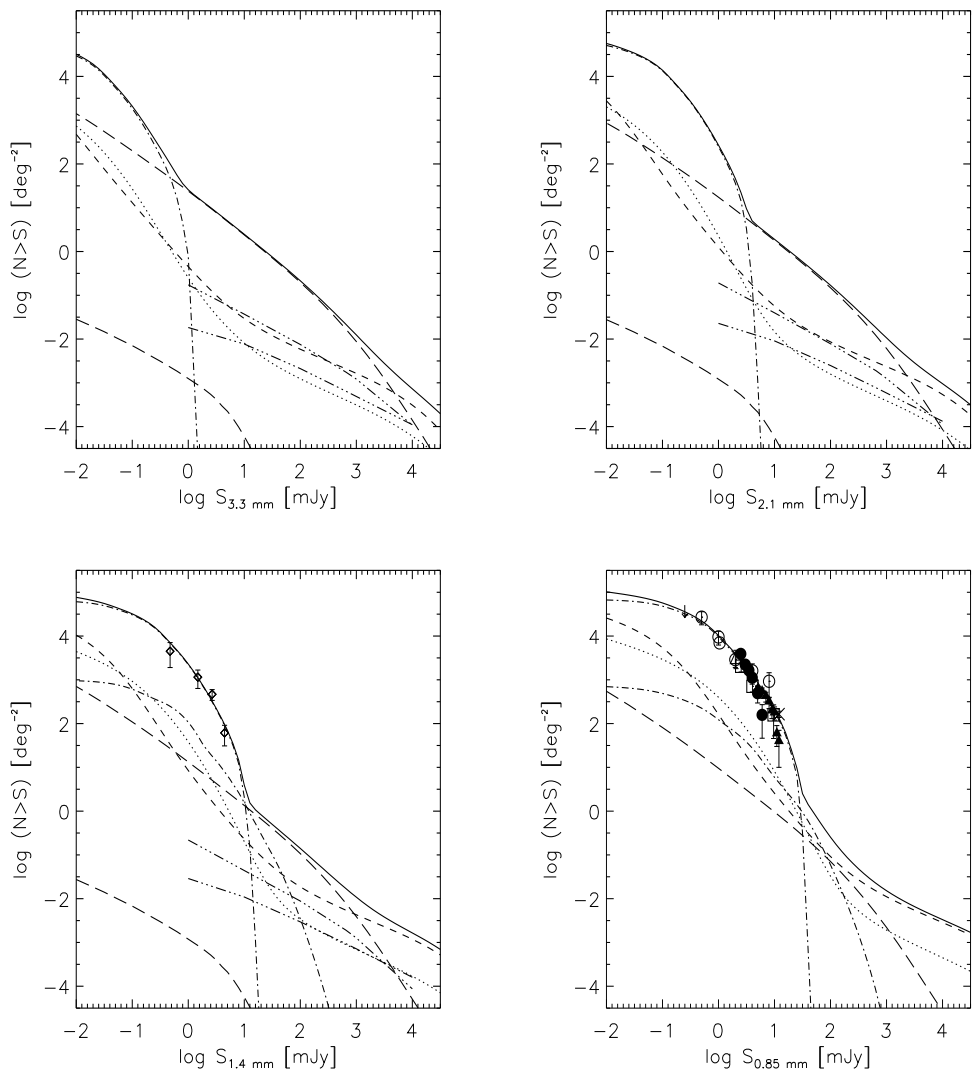}
\caption{Expected integral counts of extragalactic sources at mm
wavelengths. The solid lines show the sum of contributions from
the various populations characterized by either dust emission
(forming massive spheroidal galaxies: dot-dashed lines; spiral
galaxies: short dashes; starburst galaxies: dotted) or by
non-thermal synchrotron emission (standard steep- and
flat-spectrum radio-galaxies: upper long-dashed lines; GPS sources
with rest frame peak frequency, $\nu_{p,0}$, in the range 1--10
GHz, more numerous at fainter flux densities, and with $\nu_{p,0}
> 10\,$GHz: three dots-dashes; gamma-ray bursts: low dashed line,
not shown at 850$\,\mu$m). At 1.4 mm and $850\,\mu$m only we also
show (second set of dot-dashed lines, lower at low flux densities
but taking over at the brightest ones) the counts of strongly
lensed forming spheroids. The data points are described in
\cite{Perrotta2002b}.
 }\label{mmcounts}
\end{figure}

\section{Counts of extragalactic sources in the mm band}

Our current best guess for integral counts at some mm wavelengths
is shown in Fig.~\ref{mmcounts}. At the brightest flux densities
counts are dominated by radio sources. The counts shown here come
from the model by \cite{Toffolatti1998} which reproduces quite
well the statistics of sources at frequencies up to 8.4 GHz (see
also \cite{Toffolatti1999}). Extrapolations to higher frequencies
are liable to several sources of uncertainties: poorly known high
frequency spectra, variability, appearance of classes of sources
with poorly known space densities or whose very existence is not
clearly established.

The inflection point, occurring at $\sim 1$mJy for $\lambda =
3.3\,$mm and shifting to brighter (fainter) flux densities with
decreasing  (increasing) wavelength, signals the stepping in of a
different population: dusty galaxies, characterized by very steep
mm/sub-mm counts. Below $\sim 0.1$mJy counts are well constrained
by SCUBA and MAMBO surveys at $850\,\mu$m and 1.2 mm, respectively
(Fig.~\ref{mmcounts}). Above this flux density, we must resort to
model predictions, which show a considerable latitude. Most
current models actually predict much smoother counts in the
transition region than those shown here. However, the model
adopted here \cite{Granato2001} appears to be, at the moment, the
most astrophysically grounded. On the other hand, as discussed
below, perhaps the transition is really smoother after all,
although not for the reasons inherent in competing
phenomenological models, but because of gravitational lensing,
which introduces a significant tail, extending to $\sim 10$ times
brighter fluxes, of the counts of forming spheroidal galaxies,
which, according to \cite{Granato2001}, dominate the SCUBA/MAMBO
counts.

\begin{figure*}
\begin{center}
\epsfig{file=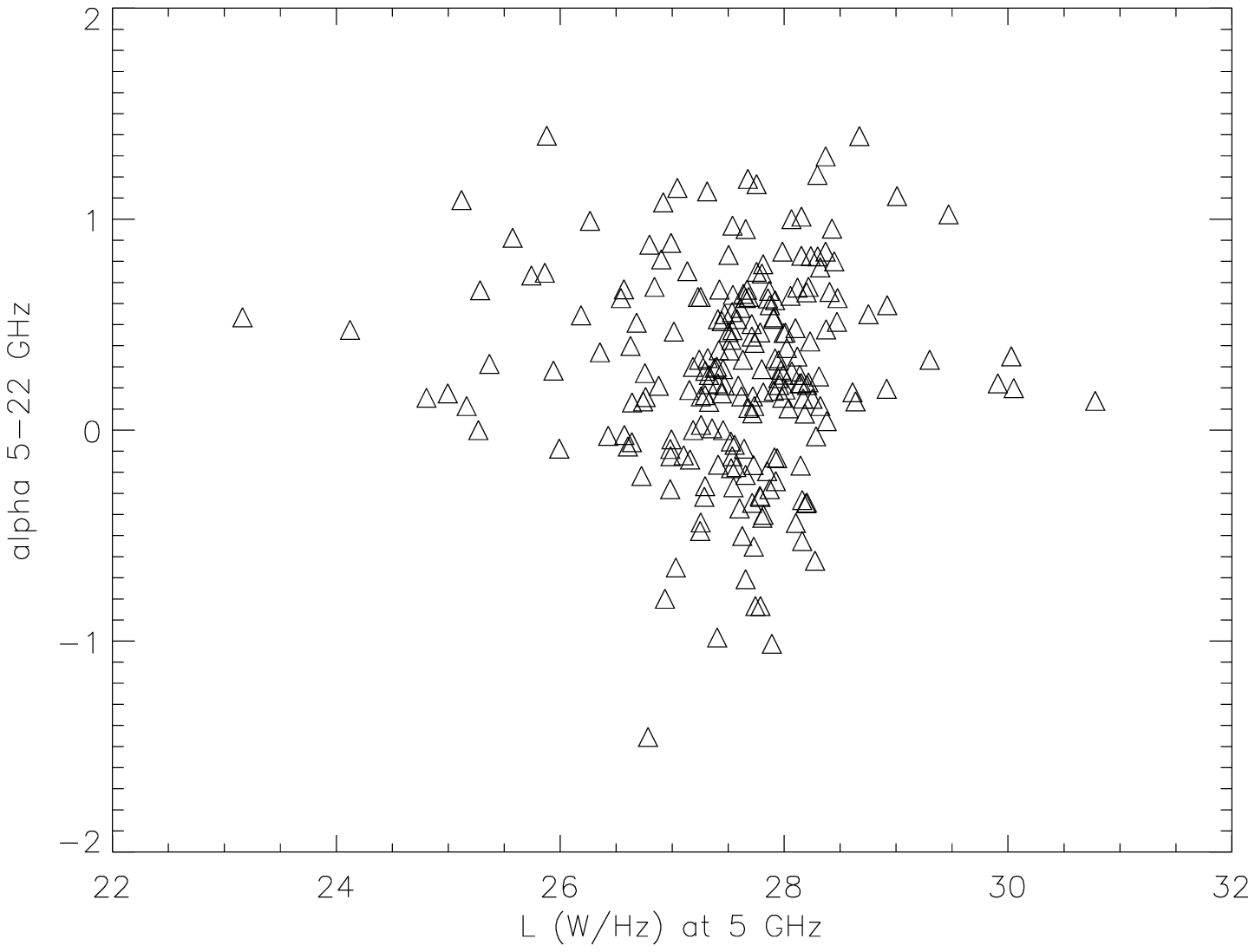,height=2.5in,width=2.5in,angle=0}
\epsfig{file=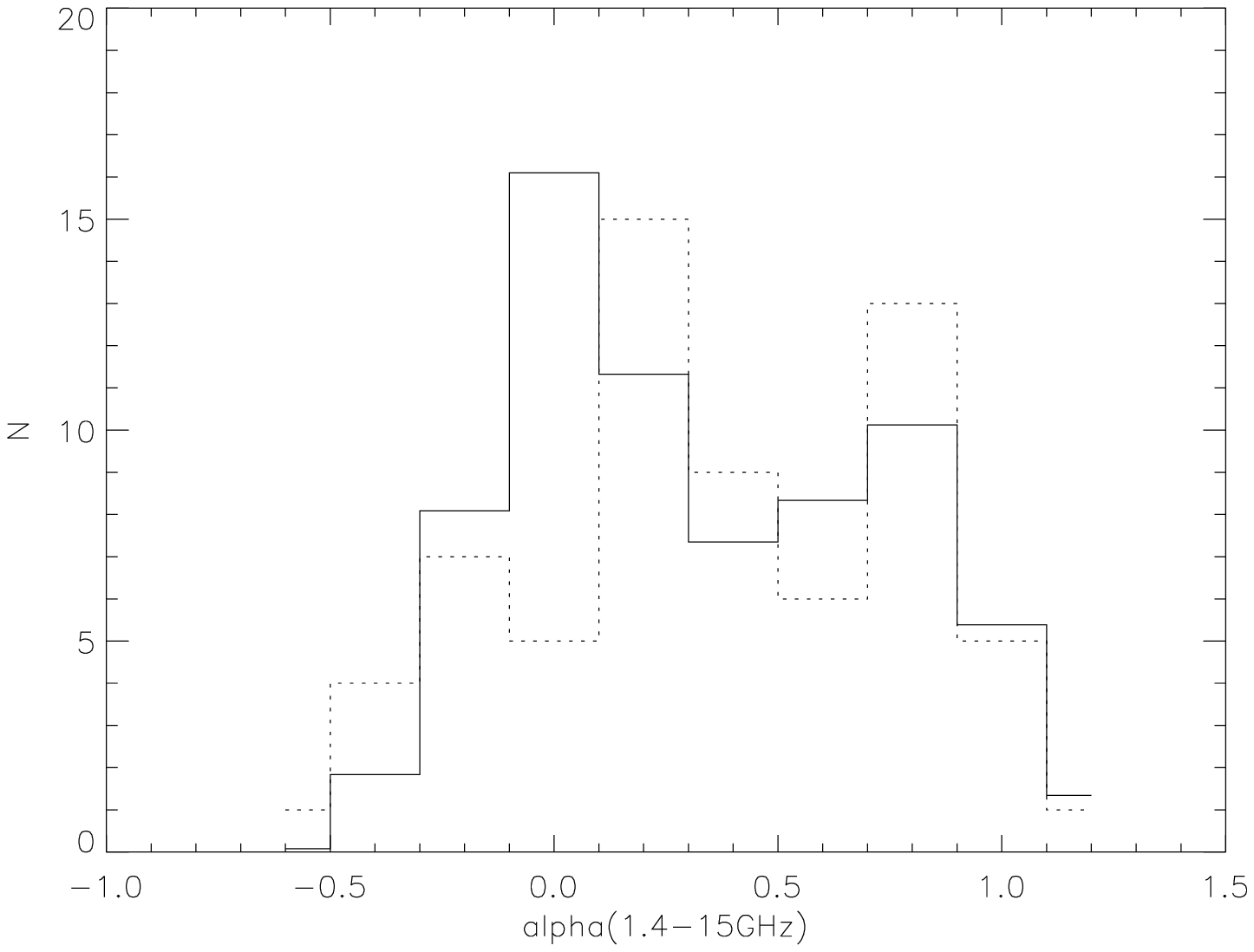,height=2.5in,width=2.5in,angle=0}
\caption{Observed (dotted) and fitted (solid) distribution of
1.4--$15\,$GHz spectral indices for the sample by
\cite{Taylor2001} (right-hand panel) and 5--22 GHz spectral index
distribution of ``flat''-spectrum ($\alpha_{1.4-5{\rm GHz}} <
0.5$) sources in K\"uhr's \cite{Kuhr1981} sample for which 22 GHz
measurements are available (left-hand panel).} \label{spinddistr}
\end{center}
\end{figure*}

\section{Radio sources}

\subsection{Sources with ``standard'' spectra}

Let us focus first on the radio-source component. What is it made
of? We know that most classes of sources fade away as we move to
mm wavelengths: steep-spectrum sources tend to further steepen
their spectra because of electron aging effects, but also it has
long been known that only a fraction of sources with flat or
rising spectra up to 5 GHz, keep them that way up to 90 GHz
\cite{OwenMufson1977,Witzel1978}; many of them become optically
thin, and therefore steep-spectrum, at mm wavelengths. Still,
these populations are expected to comprise quite a significant
fraction of sources detected at $\lsim 1\,$cm so that surveys in
this region will allow us to get very interesting information on
their physical properties. For example, the distribution of
spectral steepenings as a function of frequency provides
statistical information on the distribution of radiative ages of
sources and on mechanisms for injection and energy losses of
relativistic electrons; the transition frequency from optically
thick to optically thin synchrotron emission carries unique
information on properties of the relativistic electron population
and on the magnetic field strength of the most compact emitting
regions.

The preliminary results of the survey at $15.2\,$GHz with the Ryle
telescope \cite{Taylor2001} are already providing interesting
hints. Current models \cite{DunlopPeacock1990,Toffolatti1998},
that successfully account for the counts at frequencies up to
$8.4\,$GHz (see \cite{Toffolatti1999}) over-predict the number of
sources at the survey limit ($\simeq 20\,$mJy) by a factor $\simeq
1.5$--2, implying that the simple assumptions adopted for source
spectra break down. Preliminary results of a more detailed
analysis (Ricci \& De Zotti, in preparation) show that to account
for both the 15 GHz counts and the corresponding spectral index
distribution, significant constraints on spectral properties of
both steep- and flat-spectrum sources must be introduced. A
satisfactory fit is obtained by assuming an average steepening
above 5 GHz by $\langle \delta \alpha \rangle \simeq 0.35$ of
steep-spectrum sources ($\alpha_{1.4-5{\rm GHz}} > 0.5$) and, for
sources with flatter 1.4--$5\,$GHz spectra, a different
distribution of high-frequency (5--22 GHz) spectral indices above
and below $\log L_{5\,{\rm GHz}}(\hbox{W}/\hbox{Hz}) = 26.5$. The
model used in the right-hand side panel of Fig.~\ref{spinddistr}
assumes that, below this limit, the spectral index distribution of
sources with $\alpha_{1.4-5{\rm GHz}} < 0.5$ is a Gaussian with
mean 0.49 and dispersion 0.41, while above this limit the
distribution shown in the left-hand panel was used. For sources
with $\alpha_{1.4-5{\rm GHz}} > 0.5$ a Gaussian distribution of
spectral indices above 5 GHz was adopted, with mean 1.15 and
dispersion 0.5.

\subsection{Blazars}

But mm-wave surveys will have an even greater impact on the study
of the classes of sources with ``non-standard'' spectra, keeping
flat or inverted ($\alpha \leq 0$) up to very high frequencies.
The best known population of this kind are blazars.


Blazars are a composite population, with Spectral Energy
Distributions (SEDs) characterized by two broad peaks, the
synchrotron and the inverse Compton peak, which occur at widely
different frequencies. They have also a sort of bimodal
distribution of polarization properties. It is not yet clear
whether there is a continuity or an evolutionary link between the
various sub-populations, although some hints in this direction
have been found (e.g. inverse correlation between luminosity and
synchrotron peak frequency: \cite{Ghisellini1998,Fossati1998}).

With the better samples that can be obtained at mm wavelengths it
should be possible to test the currently favoured theoretical
paradigm, according to which the phenomenology of bright blazars
can be accounted for by a sequence in the source power and
intensity of the diffuse radiation field, surrounding the
relativistic jet, which determines the distribution of synchrotron
and inverse Compton peak frequencies. Millimetric data will
therefore enable us to study the structure and the physics of
radio jets.

Also, transition and extreme (mm peaking blazars), for which at
the moment only some circumstantial evidence is available, should
be discovered by mm surveys. For example, the two main discovery
techniques, radio and X-ray selection, have produced two
subclasses of BL Lac objects with different distributions of radio
to X-ray luminosity ratios, X-ray selected ones being
substantially weaker in the radio, although no evidence of a
significant population of radio-silent BL Lacs has been found. It
is not clear whether X-ray and radio selections sample the
extremes of the distribution of the BL Lac population or they are
discovering populations with different intrinsic properties.
Millimetric surveys are optimal for identifying and systematically
studying these sources, to assess whether there is continuity
among the sub-classes and to investigate whether they fit within
the framework of unified models. There are also hints that the
radio to mm spectral indices of faint flat-spectrum sources tend
to be steeper than those found for the brighter ones
\cite{Tornikoski2000,Tornikoski2001}. These indications, and their
interpretation, can only be tested and investigated with high
frequency surveys.

Other very interesting open issues are the characterization of
variability and polarization. Very compact sources are variable,
with typical timescales limited to $\delta t > d/c$, where $d$ is
the source size and $c$ is the velocity of light. In some cases
the shape of the spectrum was also found to vary, hinting at
evolutionary relationships between different classes of sources. A
particularly hot issue is intra-day variability (IDV), which is
most pronounced for flat-spectrum sources
\cite{WagnerWitzel1995,Krichbaum2001}. Since interstellar
scintillation, which is known to play a role in the observed cases
of IDV, should not be important at mm wavelengths, measurements in
this band would greatly help the physical interpretation of this
phenomenon.

The dominant emission process for powerful radio sources is
synchrotron whose radiation is intrinsically highly polarized. Yet
the observed polarization degree of most sources observed at MHz
or GHz frequencies is not higher than a few percent. The
depolarization may be due either to random alignment of the
magnetic field in the source or to differential Faraday rotation
of the emergent radiation. The latter effect scales as $\nu^{-2}$
and is therefore likely to be small at mm wavelengths. Therefore
polarimetric measurements in this spectral region would allow us
to measure the intrinsic polarization of sources and, together
with polarization measurements at lower frequencies, to determine
the Faraday depth. Note that the Doppler boosting of photon
frequencies along the jet axis, invoked to explain the extreme
brightness temperatures of compact sources, would make the
observed polarization for these sources liable to Faraday rotation
up to much higher frequencies than non Doppler-boosted sources.

It is worth noticing that thanks to Doppler boosting, very high
redshift blazars can show up at relatively bright radio fluxes. In
fact,  several $z> 4$ blazars have been observed at flux density
levels  $\gsim 50$--$100\,$mJy (\cite{Snellen2001} and references
therein). They are therefore convenient beacons for investigating
the abundance of $\sim 10^9\,{\hbox {M}}_\odot$ BHs and of their
host galaxies at universe ages of about 1 Gyr and, in general, to
explore the early phase of structure formation and the origin of
the radio phenomenon. The maximum angle between the jet and the
line of sight should be between 13.5 and 6.1 degrees, for the
likely range of Lorentz factors \cite{PadovaniUrry1992}, implying
that, for each beamed source, a few hundred objects not beamed in
our direction should exist at the same redshifts.

\subsection{Extreme GPS sources}

A class of sources that is expected to come out in mm surveys is
that of extreme GHz Peaked Spectrum (GPS) or  very high frequency
peakers. GPS sources are powerful ($\log P_{\rm 1.4\, GHz} \gsim
25\,\hbox{W}\,\hbox{Hz}^{-1}$), compact ($\lsim 1\,$kpc) radio
sources with a convex spectrum peaking at GHz frequencies. They
are identified with both galaxies and quasars. However,
unification models by which the two populations differ only by
effect of different orientations do not seem to apply in this
case. Rather, GPS galaxies and quasars appear to be unrelated
populations. They have different redshift, rest-frame frequency,
linear size and radio morphology distributions
\cite{Stanghellini1998,Stanghellinietal1999,Stanghellini2001,Snellen1999}.
Also, significantly different evolutionary properties are
indicated \cite{DeZotti2000}.

Sources peaking above $\simeq 10\,$GHz (in the observer's frame)
are strongly under-represented in the present samples because they
are relatively weak at the frequencies ($\leq 5\,$GHz) where large
area surveys are available. Still, some tens such sources are
known
\cite{Edge1996,Edge1998,Crawford1996,GraingeEdge1998,Tornikoski2000,Tornikoski2001,Guerra2002}.
Grainge \& Edge \cite{GraingeEdge1998} report the detection of 50
GPS sources with spectra still rising above 10 GHz (although few
details on the survey are given and the source list is not
provided); one of these has the emission peak above 190 GHz in the
rest frame.

It is now widely agreed that GPS sources correspond to the early
stages of the evolution of powerful radio sources, when the radio
emitting region grows and expands within the interstellar medium
of the host galaxy, before plunging in the intergalactic medium
and becoming an extended radio source
\cite{Fanti1995,Readhead1996,Begelman1996,Snellen2000}. Conclusive
evidence that these sources are young came from measurements of
propagation velocities. Velocities of up to $\simeq 0.4c$ were
measured, implying dynamical ages $\sim 10^3$ years
\cite{OwsianikConway1998,Owsianik1998,Polatidis1999,Taylor2000,Tschager2000}.
Estimates of radiative ages of small radio sources are also
consistent young ages \cite{Murgia1999}. The identification and
investigation of these sources is therefore a key element in the
study of the early evolution of radio-loud AGNs.

There is a clear anti-correlation between the peak (turnover)
frequency and the projected linear size of GPS sources (and of
compact steep-spectrum sources; \cite{Fanti1990,O'DeaBaum1997}),
suggesting that the process (probably synchrotron self-absorption,
although free-free absorption is also a possibility, cf.
\cite{Bicknell1997}) responsible for the turnover depends simply
on the source size. Although this anti-correlation does not
necessarily define the evolutionary track, a decrease of the peak
frequency as the emitting blob expands is indicated. Thus
millimeter-wave surveys may be able to detect these sources very
close to the moment (perhaps within the first years) when they
turn on.

On the other hand, it is not clear at this stage whether there is
a continuity between low-frequency peaked and very high-frequency
peaked GPS sources. The new samples provided by mm surveys and
follow-up VLBI observations of extreme GPS sources will allow us
to test these ideas. Millimetric surveys might also test the
frequency of occurrence of transitions from GPS to blazar spectra,
as observed for the quasar PKS0528+134 \cite{Zhang1994}, and allow
us to investigate the nature of processes involved.

The self-similar evolution models by \cite{Fanti1995,Begelman1996}
imply that the radio power drops as the source expands, so that
GPS's evolve into lower luminosity radio sources. With a suitable
choice of the parameters, this kind of models may account for the
observed counts, redshift and peak frequency distributions of the
currently available samples \cite{DeZotti2000}. On the other hand,
\cite{Snellen2000} support a scenario whereby the luminosity of
GPS sources {\it increases} with time (while the peak frequency
decreases) until a linear size $\simeq 1\,$kpc is reached, and
decreases subsequently. This scenario is consistent with the
observed positive correlation between $S_{\rm peak}$ and angular
size, although it should be noticed that this property refers to
the source population and does not necessarily reflect the
evolution of individual sources. The present data are insufficient
to conclusively prove/disprove either scenario. However the two
scenarios yield widely different predictions for counts of GPS
sources at mm wavelengths:  the first one implies that GPS sources
may comprise a quite significant fraction of bright ($S > 1\,$Jy)
radio sources at $\nu > 30\,$GHz, while very few bright GPS
sources are expected at high frequencies in the framework of the
second scenario. Bright large area mm surveys will therefore allow
us to clearly discriminate among them.


It is often stated in the literature that GPS sources hardly show
any variability \cite{O'Dea1998,Marecki1999}. However, systematic
studies are still lacking. The preliminary results of the study of
a complete sample, reported by \cite{Stanghellini1999}, indicate
low variability for GPS galaxies, while GPS quasars are found to
vary as strongly as compact flat-spectrum quasars. The GPS quasars
monitored by \cite{Tornikoski2001} do show strong variability.
Since we are dealing with very compact sources (the expected
linear size of sources peaking at $\sim 100\,$GHz is $\lsim
1\,$pc) the relevant timescales are of the order of months to
years.

Another interesting issue is polarization of GPS sources. Despite
the fact that synchrotron emission is intrinsically polarized up
to $\sim 75\%$, very low polarization is seen at cm wavelengths
($\sim 0.2\%$ at 6 cm;
\cite{O'Dea1990,Aller1992,Stanghellini1998}). At least in some
cases (particularly for GPS quasars), these low polarizations may
be attributed, to a large extent, to Faraday depolarization
(\cite{O'Dea1998} and references therein). Very large Rotation
Measures (RM$\gsim 1000\,\hbox{rad}\,\hbox{m}^{-2}$) have indeed
been found \cite{Kato1987,Aizu1990,Taylor1992,Inoue1995}. Peck \&
Taylor \cite{PeckTaylor2000} find, for a sample of Compact
Symmetric Objects, limits on the polarization degree at 8.4 GHz
with a milli-arcsec (mas) resolution of less than $1\%$. In order
to depolarize the synchrotron emission to the observed level,
Faraday RMs must reach really extreme values ($\gsim 5\times
10^5\,\hbox{rad}\,\hbox{m}^{-2}$) or, alternatively, the magnetic
field in the circumnuclear region has to be tangled on scales
smaller than $1\,$mas to produce gradients of
$1000\,\hbox{rad}\,\hbox{m}^{-2} \,\hbox{mas}^{-1}$ or more. There
are also GPS sources with small measured values of RM.

Similarly large values of RM ($\gsim
1000\,\hbox{rad}\,\hbox{m}^{-2}$) are found for only another class
of extragalactic radio sources, namely radio galaxies at the
centers of cluster cooling flows
\cite{Perley1990,Taylor1992,GeOwen1994}. This raises the question
of whether GPS sources with high RM are also in cooling flow
clusters or whether the high RMs are produced in some other way.


\subsection{Low radiative efficiency accretion onto super-massive black holes}

Millimeter-wave observations are also crucial to investigate late
stages of the AGN evolution, characterized by low radiation
efficiency. This matter was recently brought into sharper focus by
the discovery of ubiquitous, moderate luminosity
($10^{40}$--$10^{42}\,\hbox{erg}\,\hbox{s}^{-1}$) hard X-ray
emission from nearby ellipticals. VLA studies at high radio
frequencies (up to 43 GHz) have shown, albeit for a limited sample
of objects, that all of the observed compact cores of elliptical
and S0 galaxies have spectra rising up to $\simeq 20$--$30\,$GHz
\cite{DiMatteo1999}.

There is growing evidence that essentially all massive ellipticals
host black holes with masses $10^8$--$10^{10}\,$M$_\odot$ (see,
e.g., \cite{KormendyGebhardt2001}). Yet, nuclear activity is not
observed at the level expected from Bondi's \cite{Bondi1952}
spherical accretion theory, in the presence of extensive hot
gaseous halos, and for the usually assumed radiative efficiency
$\sim 10\%$ \cite{DiMatteo1999}. However, as proposed by
\cite{Rees1982}, the final stages of accretion in elliptical
galaxies may occur via rotating accretion flows
(Advection-Dominated Accretion Flows, ADAFs), characterized by a
very low radiative efficiency \cite{FabianRees1995}. The ADAF
scenario implies strongly self-absorbed thermal cyclo-synchrotron
emission due to a near equipartition magnetic field in the inner
parts of the accretion flows, most easily detected at cm to mm
wavelengths.

However the ADAF scenario is not the only possible explanation of
the data, and is not problem-free. Chandra observations of Sgr A,
at the Galactic Center, are suggestive of a considerably lower,
compared to Bondi's, accretion rate \cite{Baganoff2002}, so that
the very low ADAF radiative efficiency may not be required. Also
\cite{DiMatteo1999,DiMatteo2001} found that the high frequency
nuclear radio emission of a number of nearby early-type galaxies
is substantially below the predictions of standard ADAF models.
They conclude that, to reconcile the advection-dominated scenario
with observations, the radio emission from the inner regions must
be strongly suppressed, due perhaps to outflows or jets that dump
significant amounts of energy into the medium close to the
accretion radius, thereby reducing the accretion rates. In fact,
it was pointed out by \cite{BlandfordBegelman1999,NarayanYi1994}
that when the radiative efficiency is very low, a large fraction
of the plasma in an advection-dominated inflow/outflow solution
(ADIOS) may be unbound, leading to significant winds. Both the
intensity and the peak of the radio emission depend on the mass
loss rate. The data on sources observed so far indicate peak
frequencies in the mm region. Therefore, measurements in this
region allow very effective tests of the physics of
low-radiative-efficiency accretion.

Perna \& Di Matteo \cite{PernaDiMatteo2000} estimated that at
$\simeq 30\,$GHz the counts of advection-dominated sources can be
comparable to the total counts of radio sources estimated by
\cite{Toffolatti1998} and may be up to a factor of 10 higher if a
large fraction ($F \gsim 50\%$) of such sources have the radio
luminosity predicted by standard ADAF models, without outflows.
Although the counts by \cite{Taylor2001} already rule out large
values of $F$, much tighter constraints will come from  counts in
the millimeter region, which are even more critically dependent on
the fraction of standard ADAF sources since the emission peak
shifts from wavelengths $\lsim 1\,$mm to wavelengths $\gsim 1\,$cm
when we move from standard ADAF models (no outflow) to models with
strong outflows.

The radio emission was found to be correlated with the mass of the
central BH \cite{Franceschini1998,Salucci1999}; thus a millimeter
survey would also provide an estimate of the BH mass function.
Improved estimates can be obtained combining radio and X-ray
measurements \cite{YiBoughn1998}.

\subsection{AGNs with high frequency free-free self absorption
cutoffs.}

Free-free absorption cutoffs at frequencies $> 10\,$GHz are
expected, in the framework of the standard torus scenario for type
1 and type 2 AGNs, for radio cores seen edge on, and may have been
observed in some cases \cite{BarvainisLonsdale1998}. Again, high
radio frequency observations are essential for a comprehensive
investigation on these sources.

\subsection{Radio afterglows of gamma-ray bursts (GRBs)}

The afterglow emission of GRBs can be modelled as synchrotron
emission from a decelerating blast wave in an ambient medium,
plausibly the interstellar medium of the host galaxy
\cite{Waxman1997,WijersGalama1999,Meszaros1999}. The radio flux,
above the self-absorption break occurring at $\lsim 5\,$GHz, is
proportional to $\nu^{1/3}$ up to a peak frequency that decreases
with time. The counts of GRB afterglows at various frequencies
have been estimated by \cite{CiardiLoeb2000} (see also
\cite{SeatonPartridge2001}). As illustrated by
Fig.~\ref{mmcounts}, a large area ($>10^3\,\hbox{deg}^2$)
millimetric survey to a flux limit $\lsim 1\,$mJy might discover
some GRB. Although, according to \cite{CiardiLoeb2000}, GRB counts
would be substantially higher at cm wavelengths, higher
frequencies are more favorable to pick up the early phases of the
GRB evolution.

\section{Dusty galaxies}

As we move to fainter flux densities, the counts are dominated by
a very different population, i.e. galaxies with extreme star
formation activity, up to $10^3\,M_\odot/$year, and a different
emission process, namely dust re-radiation. It is now clear that a
large fraction, or even most, of star formation, is heavily
obscured by dust, so that mm/sub-mm surveys provide crucial pieces
of information on the early evolution of galaxies.

Furthermore, surveys at these wavelengths are optimally suited for
studying the early evolutionary stages of galaxies, as the result
of a number of concurrent factors. First, at mm wavelengths local
galaxies are very dim; this is true even for star-burst galaxies
and even more for normal spirals (the radio emission is
proportional to the star-formation rate) and for radio-quiet
ellipticals. Second, spheroidal components of galaxies must have
strongly evolved in this band: their dust emission must have been
far higher during the phases when they formed their stars. Third,
the effect of the strong luminosity evolution is further boosted
by the effect of the strongly negative K-correction.

These effects concur to produce the very steep counts, determined
by SCUBA and, more recently, by MAMBO (the Max-Planck mm bolometer
array at the IRAM 30m telescope) surveys. Surveys at relatively
longer wavelengths (mm vs sub-mm) are advantageous for picking up
very high redshift forming galaxies, whose observed dust emission
at sub-mm wavelengths may come from frequencies beyond the dust
emission peak.

Understanding the physical and evolutionary properties of these
sources is still a major challenge. Even the best semi-analytic
models \cite{Cole2000,DevriendtGuiderdoni2000} hinging upon the
standard picture for structure formation in the framework of the
hierarchical clustering paradigm, tuned to agree with detailed
numerical simulations, are stubbornly unable to account for the
(sub)-mm counts of galaxies. In fact, the canonical hierarchical
clustering models for the formation of galaxies rather predicted
that most star formation in the universe occurred within
relatively small proto-galaxies, at typical rates of
$10\,M_\odot/$year, that later merged to form larger galaxies. The
data are more consistent with the traditional ``monolithic''
approach whereby giant ellipticals were already present at
substantial redshifts ($z\gsim 2$), and formed most of their stars
in a single gigantic starburst, followed by essentially passive
evolution. The ``monolithic'' approach, however, is inadequate to
the extent that it cannot be fitted in a consistent scenario for
structure formation from primordial density fluctuations.

A possible way out was proposed by \cite{Granato2001} who
elaborated a scheme whereby the early evolution of giant
ellipticals is tightly inter-related with that of quasars (see
also \cite{EalesEdmunds1996,SilkRees1998}), and feed-back effects,
from supernova explosions and from active nuclei, delay the
collapse of baryons in smaller clumps while large ellipticals form
their stars as soon as their potential wells are in place, so that
the canonical hierarchical CDM scheme -- small clumps collapse
first -- is reversed for baryons. The counts predicted by this
model are shown and compared with SCUBA and MAMBO data in
Fig.~\ref{mmcounts}.

This scheme entails a number quantitative, testable predictions,
in particular about the clustering and gravitational lensing of
forming spheroidal galaxies. SCUBA/MAMBO galaxies are expected to
be highly biased tracers of the dark matter distribution and
therefore strongly clustered. Their clustering properties are
indicative of their halo masses. Recent results by
\cite{Magliocchetti2001,Perrotta2002b} show that the model
predictions are nicely consistent with the (still limited) data.
The estimated $w(\theta)$ for bright SCUBA galaxies
\cite{Scott2002} and for EROs \cite{Daddi2002,Daddi2000,Daddi2001}
indicate dark halo masses $\sim 10^{13}\,\hbox{M}_\odot$ for these
sources; the $w(\theta)$ for LBGs \cite{Giavalisco1998} indicate
typical masses at least 10 times lower (see also
\cite{MoustakasSomerville2002}).

Also, the bright tail of mm/sub-mm counts of forming spheroids is
predicted to be strongly affected by gravitational lensing. In
fact, although the probability of strong lensing is very small, it
has  a power-law tail ($p(A) \propto A^{-3}$) extending up to
large values. Thus,  if counts are steep enough, the fraction of
lensed sources at bright fluxes may be large. The mm/sub-mm region
is ideally suited for detecting them because, on one side, high
redshift sources (with highest probability of being
gravitationally lensed) are greatly emphasized by steep, negative
K-correction plus evolution, and, on the other side, the model by
\cite{Granato2001} predicts extremely steep counts, reflecting the
exponential decline of the Press-Schechter mass function (there
are no low-$z$ counterparts to forming spheroids: their formation
is essentially completed by $z\simeq 2$). Accurate modelling of
the lensing effect on counts \cite{Perrotta2002a} shows that
almost all spheroids detected at $850\,\mu$m flux densities larger
than $\sim 60\,$mJy are strongly amplified by gravitational
lensing. Even accounting for all the contributing populations, the
fraction of lensed objects at $S_{850\mu{\rm m}} \sim 70\,$mJy
should remain at the level of about $40\%$ \cite{Perrotta2002b}.

\section{Sunyaev-Zeldovich (SZ) effect from quasar-driven blast waves}

We end in a somewhat more speculative vein, introducing a
population of sources that may really be yet to be discovered. It
is natural to expect that extremely powerful sources, such as
quasars, strongly affect the surrounding medium (e.g., they may
photo-ionize the intergalactic medium (IGM) \cite{ReesSetti1970}).
The idea of strong shock heating of the medium by energetic
outflows from early quasars, originally developed by
\cite{Ikeuchi1981}, has been recently revived by
\cite{Natarajan1998,NatarajanSigurdsson1999,Aghanim2000}. The main
weakness of these analyses is the lack of a convincing physical
mechanism or of empirical evidence that a significant fraction of
the power generated by quasars comes out as supersonic winds
(except in the case of BAL quasars) and goes into heating of the
 medium. Recent analyses suggest, however, that quasars
are the best candidates for accounting for the pre-heating of gas
in groups and clusters of galaxies \cite{Balogh2001,Bower2001}: it
is enough that 1--5\% of the bolometric emission of the quasars
goes into heating of the IGM. But then \cite{Platania2002} the
heated gas within the host galaxy produces a SZ effect of
amplitude (in the Rayleigh-Jeans region)
$$\left|\left({\Delta T \over T}\right)_{\rm RJ}\right|  =  {1\over
2} {\Delta \epsilon \over \epsilon_{\rm CMB}} \nonumber $$
with $\Delta \epsilon = f_h (E_{\rm bol}/V) (t_{\rm bw}/t_c)$,
where $f_h$ is the fraction of the total energy ($E_{\rm bol}$)
emitted by the quasar  going into heating of the ambient medium,
$t_{\rm bw}$ is the lifetime of the blast-wave, $V$ is the volume
occupied by the heated gas, and $t_c$ is its Compton cooling time.
For $t_{\rm bw} \simeq t_{\rm exp}$ ($t_{\rm exp}$ being the
expansion timescale) we have
$$\left|\left({\Delta T \over T}\right)_{\rm RJ}\right| \simeq 3.4
\times 10^{-4} {f_h\over 0.01} {E_{\rm bol}\over
10^{62}\hbox{erg}\,\hbox{s}^{-1}} {\hbox{Mpc}^3\over V} {50 \over
H_0} (1+z)^{-3/2}\nonumber$$
on an angular scale $\theta \sim 100" R(\hbox{Mpc})$ for $z\gsim
1$ ($R\simeq v_{\rm bw}t_{\rm bw}$, $v_{\rm bw}$ being the outward
velocity of the shocked shell). These figures are not far from the
results reported by Richards et al. (1997): $|\delta T/T| \sim
10^{-4}$ over an area of $30''\times 65"$, and by Jones et al.
(1997): $|\delta T/T| \sim 1.4\times 10^{-4}$ in a beam of
$100''\times 175"$. Estimates of the counts of such SZ effects
have been worked out by \cite{Platania2002} who also pointed out
that, due to the relatively small angular scale of these signals,
they may be swamped by radio and far-IR emissions of galaxies
hosting the quasars. Therefore, to detect them, we should take
advantage of the frequency region where these sources are weaker,
i.e., once again, of the mm to cm range.

\section{Conclusions}

Sub-millimeter (SCUBA) and millimeter (MAMBO) surveys have played
a crucial role in shaping our understanding of early phases of
galaxy formation. In fact, they have proven that most of the
star-light at high redshifts ($z \gsim 2$) has been reprocessed by
dust so that the optical view of the cosmic star-formation history
is highly incomplete and biased. On the other hand, such surveys
have covered only small areas: from $\sim 0.1\,\hbox{deg}^2$ for
the shallowest ones (flux limit $\lsim 10\,$mJy;
\cite{Scott2002,Borys2002,Carilli2000,Bertoldi2000}), to $\sim
0.01\,\hbox{deg}^2$ for the deep ones
\cite{Smail1997,Hughes1998,Blain1999,Barger1999}. Redshift
estimates (e.g. \cite{Dunlop2001}) indicate that many (most?) of
the detected sources are likely at substantial redshifts ($z \gsim
2$) and correspond to extreme star-formation activity (up to $>
10^3\,\hbox{M}_\odot\,\hbox{yr}^{-1}$). Much more extensive
surveys are necessary to adequately sample the mm/sub-mm
luminosity function of star-forming galaxies. Larger area surveys
would also allow us to assess the clustering properties of these
sources \cite{Magliocchetti2001,Perrotta2002b}, which provide
unique information on masses of the associated dark halos and,
correspondingly, crucial tests for scenarios for the formation of
structures. As discussed by \cite{Perrotta2002a,Perrotta2002b},
quite a significant number of high-redshift forming spheroidal
galaxies are expected at relatively bright fluxes (above several
tens of mJy at $\lambda \sim 1\,$mm) due to strong gravitational
lensing. Detection of these sources would offer a precious
opportunity for investigating their properties and to learn about
the distribution of matter in the high-$z$ universe.

At somewhat longer wavelengths radio galaxies play an increasingly
important role. Surveys at mm to few-cm wavelengths will provide
key information on the physical properties of ``standard'' steep-
and flat-spectrum sources and will allow comprehensive
investigations of rare, but very interesting, populations of
sources with ``non-standard'' spectra, rising with increasing
frequency or keeping flat up to very high frequencies. Such
classes of sources are very difficult to single out at low
frequencies where they are swamped by the much more numerous
``standard'' populations. Important examples are blazars, GPS
sources, and ADAF/ADIOS sources. Among the many very interesting
issues on blazars that can be addressed there are the
relationships among the different sub-classes (high- and
low-polarization, high- and low-frequency peaked BL Lacs) and the
understanding of their variability properties, in particular of
intra-day variability. Also, thanks to their strong Doppler
boosting, blazars can be seen at relatively bright flux densities
up to very high redshifts, thus offering a precious opportunity to
investigate the presence of super-massive black holes at early
epochs and the origin of the radio phenomenon.

Both the earliest and the latest phases of nuclear activity show
up most clearly at mm wavelengths. On one side, extreme GPS
sources, whose spectra peak at mm wavelengths, probably correspond
to the very earliest phases of the evolution of radio sources
(perhaps within the first years). Current evolutionary scenarios
predict widely different luminosities for these sources, so that
mm surveys can easily discriminate among them.

Also the latest phases of the evolution of active nuclei,
characterized by low accretion rates and/or low radiative
efficiency (ADAF/ADIOS models), show inverted radio spectra, with
turnover frequencies in the millimeter range. According to some
models the turnover frequency is related to the rate of mass loss
on outflows. The counts of such sources at mm wavelengths are
highly sensitive to the accretion/outflow rates.

Yet another class of extragalactic sources with inverted radio
emission spectrum are afterglows of gamma-ray bursts. Although
they are generally quite faint, some may be discovered by mm
surveys covering areas $\gsim 10^3\,\hbox{deg}^3$ to a flux limit
$\lsim 1\,$mJy. Again, the mm selection picks up very early phases
of their evolution.

The above are just examples. It is quite likely that, in addition
to the many other areas of scientific interest that may be added,
unexpected phenomena will show up as this band opens  to surveys.

\acknowledgments LT acknowledges partial financial support from
the European Community, Research Training Network Programme,
Contract n.HPRN--CT--2000--00124. Work supported in part by MIUR
and ASI.

%
%

\end{document}